\documentclass{aa}
\usepackage{graphics,latexsym,natbib}
\usepackage{times,graphicx,epsfig}
\begin{document}
\title{The asymmetric drift, the local standard of rest, and implications from RAVE data}

\author{O.~Golubov$^{1,2,3}$,
A. Just$^{1}$, O. Bienaym\'{e}$^{4}$, J. Bland-Hawthorn$^{5}$, B. K. Gibson$^{6,7,8}$,
E. K. Grebel$^{1}$, U. Munari$^{9}$, J. F. Navarro$^{10}$, Q. Parker$^{11,12,13}$, G. Seabroke$^{14}$, W. Reid$^{11,12}$,
A. Siviero$^{15}$, M. Steinmetz$^{16}$, M. Williams$^{16}$, F. Watson$^{13}$, and T. Zwitter$^{17,18}$}

\offprints{A. Just}
\mail{just@ari.uni-heidelberg.de}
\institute{$^{1}$Astronomisches Rechen-Institut, Zentrum f\"{u}r Astronomie der Universit\"{a}t Heidelberg, M\"{o}nchhofstr. 12--14, Heidelberg 69120, Germany\\
$^{2}$Department of Aerospace Engineering Sciences, University of Colorado at Boulder, 429 UCB, Boulder, CO, 80309, USA\\
$^{3}$Institute of Astronomy, Kharkiv National University, 35 Sumska Str., Kharkiv, 61022, Ukraine\\
$^{4}$Observatoire astronomique de Strasbourg, 11 rue de l'Universit\'{e}, Strasbourg, France\\
$^{5}$Sydney Institute for Astronomy, School of Physics A28, University of Sydney, NSW 2006, Australia\\
$^{6}$Dept of Phys \& Astro, Saint Mary’s Univ, Halifax, B3H 3C3, Canada\\
$^{7}$Monash  Centre for Astrophysics, Clayton, 3800, Australia\\
$^{8}$Jeremiah Horrocks Institute, UCLan, Preston, PR1 2HE, UK\\
$^{9}$NAF Osservatorio Astronomico di Padova, Asiago I-36012, Italy\\
$^{10}$Department of Physics \& Astronomy, University of Victoria, Victoria, BC, Canada V8P 5C2\\
$^{11}$Department of Physics \& Astronomy, Macquarie University, Sydney, NSW 2109, Australia\\
$^{12}$Macquarie Research Centre for Astronomy, Astrophysics and Astrophotonics\\
$^{13}$Australian Astronomical Observatory, PO Box 296, Epping, NSW 2121, Australia\\
$^{14}$Mullard Space Science Laboratory, University College London, Holmbury St Mary, Dorking, RH5 6NT, UK\\
$^{15}$Department of Physics and Astronomy, Padova University, Vicolo dell'Osservatorio 2, I-35122 Padova, Italy\\
$^{16}$Leibniz-Institut f\"{u}r Astrophysik Potsdam (AIP), An der Sternwarte 16, D-14482 Potsdam, Germany\\
$^{17}$Faculty of Mathematics and Physics, University of Ljubljana, Jadranska 19, SI-1000 Ljubljana, Slovenia\\
$^{18}$Center of Excellence SPACE-SI, Askerceva cesta 12, SI-1000 Ljubljana, Slovenia}

\date{Printed: \today}


  \abstract
{The determination of the local standard of rest (LSR), which corresponds to the measurement of the peculiar motion of the Sun based on the derivation of the asymmetric drift of stellar populations, is still a matter of debate. The classical value of the tangential peculiar motion of the Sun with respect to the LSR was challenged in recent years, claiming a significantly larger value.}
{We present an improved Jeans analysis, which allows a better interpretation of the measured kinematics of stellar populations in the Milky Way disc. We show that the RAdial Velocity Experiment (RAVE) sample of dwarf stars is an excellent data set to derive tighter boundary conditions to chemodynamical evolution models of the extended solar neighbourhood.}
{We propose an improved version of the Str\"omberg relation with the radial scalelengths as the only unknown.  We redetermine the asymmetric drift and the LSR for dwarf stars based on RAVE data. Additionally, we discuss the impact of adopting a different LSR value on the individual scalelengths of the subpopulations.}
{Binning RAVE stars in metallicity reveals a bigger asymmetric drift
(corresponding to a smaller radial scalelength) for more
metal-rich populations. With the standard assumption of velocity-dispersion independent radial scalelengths in each metallicity bin, we redetermine the LSR.
The new Str\"omberg equation yields a joint LSR value of $V_\mathrm{\sun}=3.06\pm0.68$ km s$^{-1}$,
which is even smaller than the classical value based on Hipparcos data. The corresponding radial scalelength increases from 1.6 kpc for the metal-rich bin to 2.9 kpc for the metal-poor bin, with a trend of an even larger 
scalelength for young metal-poor stars.
When adopting the recent Sch\"onrich value of $V_\mathrm{\sun}=12.24$ km s$^{-1}$ for the LSR, the new Str\"omberg equation yields much larger individual radial scalelengths of the RAVE subpopulations, which seem unphysical in part.}
{The new Str\"omberg equation allows a cleaner interpretation of the kinematic data of disc stars in terms of radial scalelengths. Lifting the LSR value by a few km s$^{-1}$ compared to the classical value results in strongly increased radial scalelengths with a trend of smaller values for larger velocity dispersions.}


\titlerunning{Asymmetric drift and LSR}
\authorrunning{Golubov, Just, et al.}

\maketitle

\section{Introduction}

In any dynamical model of the Milky Way, the rotation curve (which is the circular speed $v_\mathrm{c}(R)$ as function of distance $R$ to the Galactic centre) plays a fundamental role. In axisymmetric models the mean tangential speed $v_\mathrm{c}$ of stellar subpopulations deviates from $v_\mathrm{c}$, which is quantified by the asymmetric drift $V_\mathrm{a}$. Converting observed kinematic data (with respect to the Sun) to a Galactic coordinate system requires additionally the knowledge of the peculiar motion of the Sun with respect to the local circular speed.

The asymmetric drift of a stellar population is defined as the difference between the velocity of a hypothetical set of stars
possessing perfectly circular orbits and the mean rotation velocity of the population under consideration.
The velocity of the former is called the standard of rest.
If the measurements are made at the solar Galactocentric radius, it is the local standard of rest, or LSR.
The determination of the LSR corresponds to measuring the peculiar motion $(U_\odot,V_\odot,W_\odot)$ of the Sun,
where $U_\odot$ is the velocity of the Sun in the direction of the Galactic centre, $V_\odot$ in the direction of the Galactic rotation,
and $W_\odot$ in the vertical direction.
While measuring $U_\odot$ and $W_\odot$ is relatively straightforward,
$V_\odot$ requires a sophisticated asymmetric drift correction for its measurement,
which is one goal of this paper.
The asymmetric drift  $V_\mathrm{a}=v_\mathrm{c}-\overline{v_\mathrm{\phi}}=\Delta V - V_\mathrm{\sun}$ is the difference of the local circular speed $v_\mathrm{c}$ and the mean rotational speed $\overline{v_\mathrm{\phi}}$ of the stellar population. The asymmetric drift corresponds (traditionally with a minus sign to yield positive values for $V_\mathrm{a}$) to the measured mean rotational velocity of the stellar
sample corrected by the reflex motion of the Sun.

The main problem is to disentangle the asymmetric drift $V_\mathrm{a}$ of each subpopulation and the peculiar motion of the Sun $V_\odot$ using measured mean tangential velocities $(\overline{v_\mathrm{\phi}}-v_\odot) = -\Delta V=-(V_\mathrm{a} + V_\mathrm{\sun})$. For any stellar subpopulation in dynamical equilibrium, the Jeans equation (Eq.~\ref{jeans}) provides a connection of the asymmetric drift, radial scalelengths, and properties of the velocity dispersion ellipsoid in axisymmetric systems \citep{binney}. There are two principal ways to determine both $V_\mathrm{a}$ and $V_\odot$ with the help of the Jeans equation. The direct path would be to measure for one tracer population the radial gradient of the volume density $\nu$ and of the radial velocity dispersion in the Galactic plane together with the inclination of the velocity ellipsoid away from the Galactic plane additionally to the local velocity ellipsoid. This approach is still very challenging due to observational biases in spatially extended stellar samples (by extinction close to the midplane, distance-dependent selection biases etc.). Therefore we need to stick to the classical approach to apply the Jeans equation to a set of subpopulations and assume common properties or dependencies of the radial scalelengths and the velocity ellipsoid.

On top of this basic equilibrium model, non-axisymmetric perturbations like spiral arms may lead to a significant shift in the local mean velocities of tracer populations \citep[see][for the first direct measurement of a gradient in the mean radial velocity and the interpretation in terms of spiral arms]{siebert11a, siebert}. In the present paper we focus on the discussion of the Jeans equation in axisymmetric models.

In the classical approach the Jeans equation is applied to local stellar samples of different (mean) age, which show increasing velocity dispersion with increasing age due to the age-velocity dispersion relation. Up to the end of the last century, the general observation that the mean tangential velocity depends linearly on the squared velocity dispersion for stellar populations that are not too young allowed the measurement of $V_\odot$ by extrapolation to zero velocity dispersion. The corresponding reformulation of the Jeans equation is the famous linear Str\"omberg equation (Eq.~\ref{stromberg}).
This method was also used by \citet{dehnen98} for a volume-complete sample of Hipparcos stars to constrain the LSR. They found again that the asymmetric drift $V_\mathrm{a}$ depends linearly on the squared
(three-dimensional) velocity dispersion of a stellar population. A linear
extrapolation to zero velocity dispersion led to the LSR.
The velocity of the Sun in the direction of the Galactic rotation with respect to the LSR appeared to be $V_\mathrm{\sun}=5.25\pm 0.62$ km s$^{-1}$.
\citet{aumer09} applied a similar approach to the new reduction of the Hipparcos catalogue
and obtained the same value $V_\mathrm{\sun}=5.25\pm 0.54$ km s$^{-1}$, but with
a smaller error bar.
The linear Str\"omberg relation \citep{binney} adopted in this analysis relies on the
crucial assumption that the structure
(radial scalelengths and shape of the velocity dispersion ellipsoid) of the
subpopulations with different velocity dispersions are similar.

In recent years it was argued, based on very different methods, that the value of $V_\mathrm{\sun}$ should be increased significantly.
Based on a sophisticated dynamical model of the extended solar neighbourhood, \citet{binney10a} argued that the $V$ component of the Sun's peculiar velocity should be revised upwards to $ \approx 11$ km s$^{-1}$.
In \citet{mcmillan10} it was shown that $V_\mathrm{\sun}\approx 11$ km s$^{-1}$ would be more appropriate based on the space velocities of maser sources in star-forming regions \citep{reid09}.
The chemodynamical model of the Milky-Way-like galaxy of \citet{schoenrich10} shows a non-linear dependence $V_\mathrm{a}(\sigma_\mathrm{R}^2)$. This implies different radial
scalelengths
and/or different shapes of the velocity ellipsoid for different subpopulations.
Fitting the observed dependence $V_\mathrm{a}(\sigma_\mathrm{R}^2)$ by predictions of their model,
they got $V_\mathrm{\sun}=12.24\pm 0.47$ km s$^{-1}$, which is also significantly larger than
the classical value. Most recently \citet{bovy12b} derived an even larger value of $V_\mathrm{\sun}\approx 24$ km s$^{-1}$ based on Apogee data and argued for an additional non-axisymmetric motion of the locally observed LSR of 10 km s$^{-1}$ compared to the real circular motion.

In view of the inside-out growth of galactic  discs (established by the observed radial colour and metallicity gradients), there is no a priori reason why stellar subpopulations with different velocity dispersion should have similar radial scalelengths independent of the significance of radial migration processes \citep{matteucci89, matteuchi01, wielen96, schoenrich09, scannapieco11, minchev13}. Therefore it is worthwhile to step back and investigate the consequences of the Jeans equation in a more general context. The fact that the peculiar motion of the Sun (i.e. the definition of the LSR) is one and the same unique value entering the dynamics of all stellar subpopulations already shows that changing the observed value for $V_\mathrm{\sun}$ will have a wide range of consequences for our understanding of the structure and evolution of the Milky Way disc.

The goal of this paper is twofold. We discuss the Jeans equation in a more general context and derive a new version of the Str\"omberg equation that is useful for an improved method to analyse the interrelation of radial scalelengths, the asymmetric drift, and the LSR. We emphasize the impact of different choices of LSR. Secondly, we apply the new method to the large and homogeneous sample of dwarf stars provided by the latest internal data
release (May 15th, 2012) of the RAVE \citep[see][for the first, second, and third data release respectively]{RAVE-DR1,zwitter08,RAVE-DR3} and complement it with other data sets.
In Section 2 we describe the data analysis, Section 3 contains the Jeans analysis, in Section 4 our results are presented, and Section 5 concludes with a discussion.

\section{Data analysis}
For our analysis we use several different kinematically unbiased data sets.
In all the cases, only stars with heliocentric distances $r<3$\,kpc and
Galactocentric radii 7.5 kpc$<R<$8.5 kpc
and with distances to the mid-plane $|z|<500$pc are selected.
A list of variables used in the paper are collected in Table \ref{tabvar}.

\begin{table}
\caption{Variables used in the paper.}
\label{tabvar}
\begin{tabular}{cl}
\hline
Variables & Definitions \\ \hline
\multicolumn{2}{c}{\textit{local coordinates}} \\ 
$r$ &  distance \\
$v_r$ & line-of-sight velocity \\
$\mu$ & proper motion with respect to the Sun \\
$U$, $V$, $W$ & velocities with respect to the LSR \\
$U_\mathrm{\sun}$, $V_\mathrm{\sun}$, $W_\mathrm{\sun}$ & velocity of the Sun with respect to the LSR \\
$\Delta V$ & $v_\odot-\overline{v_\mathrm{\phi}}$ \\
$V_a$ & $v_\mathrm{c}-\overline{v_{\phi}}$, asymmetric drift \\
$V'$ & Eq. \ref{Vprime} \\
\multicolumn{2}{c}{\textit{galactocentric variables}} \\
$R_0$ & Galactocentric radius of the Sun \\
$v_\mathrm{\sun}$ & velocity of the Sun around the Galactic centre \\
$v_\mathrm{c}$ & velocity of the LSR around the Galactic centre \\
$v_r$, $v_\phi$, $v_z$ & velocities in cylindrical Galactocentric coordinates \\
$\sigma_r$, $\sigma_\phi$, $\sigma_z$ & velocity dispersions \\
$R$ & Galactocentric radius \\
$z$ & height above the Galactic plane \\
\multicolumn{2}{c}{\textit{radial scalelengths}} \\
$R_E$ & kinetic energy scalelength $(R_\nu^{-1}+R_{\sigma}^{-1})^{-1}$ \\
$R_\nu$ & scalelength of the tracer density $\nu$ \\
$R_\mathrm{d}$ & scalelength of the total surface density \\
$R_{\sigma}$ & scalelength of $\sigma_R^2$  \\  \\
\end{tabular}
\end{table}

Even though most stars in our samples are relatively local, we make all computations in Galactocentric cylindrical coordinates.
That is why we need to fix the Galactocentre distance $R_0$ and the circular speed $v_\mathrm{\sun}$  for our computations:
they influence how velocities of distant stars are decomposed into radial and rotational components.
We adopt $R_0=8$\,kpc, which is
consistent with most observational data to date
\citep{reid93,gillessen09}.
Assuming Sgr A* to reside at the centre of the Galaxy at rest
and taking $\mu_{\mathrm{l,A\!^{*}}}=6.37 \pm 0.02$ mas yr$^{-1}$ for its proper motion
in the Galactic plane \citep{reid05},
we find the rotation velocity of the Sun to be $v_\mathrm{\sun} =
241.6$ km s$^{-1}$ in a Galactocentric coordinate system.
This velocity consists of the circular velocity in the solar neighbourhood $v_\mathrm{c}$  (of the LSR)
and the peculiar velocity of the Sun with respect to the LSR
$V_\mathrm{\sun}$,
so that $v_\mathrm{\sun}=v_\mathrm{c}+V_\mathrm{\sun}$.
For the radial and vertical components of the LSR, we assume
$U_\mathrm{\sun}=9.96$ km s$^{-1}$ and $W_\mathrm{\sun}=7.07$ km s$^{-1}$ from \citet{aumer09}.

Any radial or vertical gradient of the mean velocity and velocity dispersions may influence the determination of the corresponding values at the solar position. Linear trends cancel out for symmetric samples with respect to the solar position, but spatially asymmetric samples can result in shifts of mean velocity and velocity dispersions.
Additionally, spatial gradients of the mean velocities result in an overestimation of the velocity dispersions due to the shifted mean values at the individual positions of the stars. For example, for the tangential velocity dispersion $\sigma_{\phi}$ we find
\begin{equation}
\sigma_{\phi}^2 = \overline{\left(v_\mathrm{\phi}-\overline{v_{\phi}}\right)^2}
-\overline{\left(\delta v_{\phi}\right)^2}
-\left(\delta\overline{v_{\phi}} \right)^2.
\label{sigmaphi}
\end{equation}
Here $v_\mathrm{\phi}$ is the tangential velocity of the stars, $\overline{v_{\phi}}$ the sample mean, $\delta\overline{v_{\phi}}$ is the root mean square (rms) value of the difference of the mean tangential velocity at the individual positions of the stars to the sample mean, and $\delta v_{\phi}$ is the rms of the propagated individual measurement errors. In our analysis we do not take the described effects into account but discuss the potential impact on our results.

Despite the RAVE sample being the biggest one, supplementing it with other samples provides an important consistency check
as all samples have different selection criteria and biases, different sources of distance measurements,
and are differently divided into subsamples with different kinematics.

\subsection{RAVE data}

For the upcoming fourth data release the stellar parameter pipeline to derive effective temperature, surface gravity, and metallicity was improved significantly. The latest internal data release is based on the new stellar parameter pipeline and contains 402\,721 stars.
Internally, there are two independent catalogues of distances available. The first is based on isochrone fitting in the colour-magnitude diagram (CMD) \citep{zwitter10}, which contains 383\,387 stars in the updated version. The second method is based on a Bayesian analysis of the stellar parameters \citep{burnett11} and contains 201\,670 stars.
For these stars line of sight velocities $v_r$, proper motions $\mu$, temperatures $T_\mathrm{eff}$, surface gravities $\log g$, and  metallicities [M/H] are measured.
The $J$ and $K$ colours are taken from the Two Micron All Sky Survey (2MASS) \citep{2mass}.

\begin{figure}
 \includegraphics[width=85mm,angle=0]{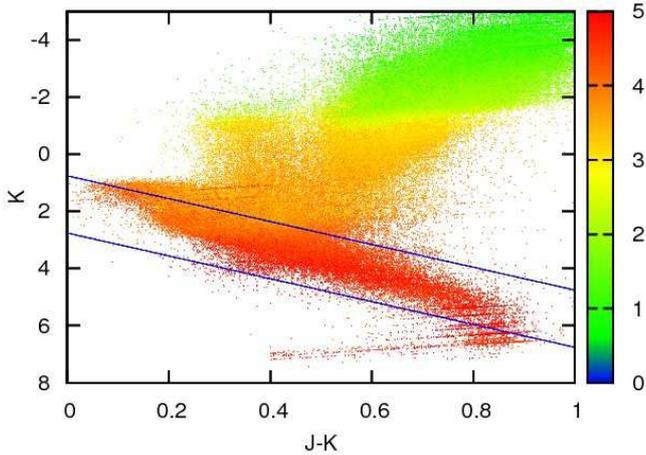}
  \caption{CMD of the full RAVE sample based on Zwitter distances. Surface gravity $\log g$ is colour-coded and the lines show the selected dwarf stars.}
\label{cmd}
\end{figure}

For our analysis we selected stars with absolute distance errors $\Delta r/r
\leq 0.3$,
proper motion errors $\Delta\mu \leq 10$ mas yr$^{-1}$, radial velocity errors $\Delta
v_r \leq 3$ km s$^{-1}$, Galactic latitudes $|b| \geq 20^\circ$.
The CMD of the selected stars is shown in Figure \ref{cmd} with colour-coded  $\log g$.
Furthermore, we selected only stars that meet the criterion $0.75<K-4(J-K)<2.75$ (see Figure \ref{cmd}),
primarily to exclude subgiants and giants.
Finally, a total number $N=68\,670$ stars remain.

\begin{figure}
  \includegraphics[height=85mm,angle=270]{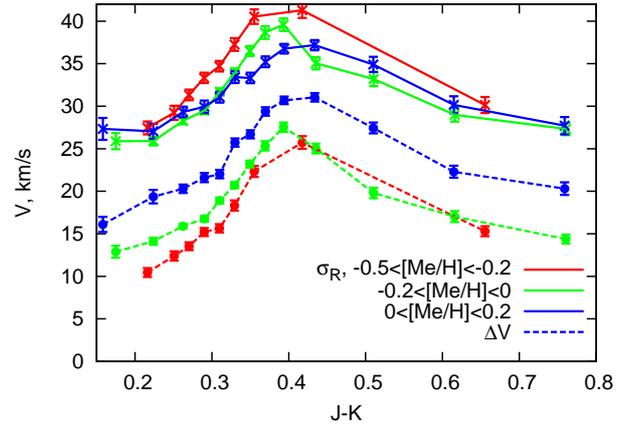}
  \caption{Measured mean tangential velocity $\Delta V$ (full circles) and radial velocity dispersion $\sigma_\mathrm{R}$ (crosses) of the RAVE sample based on Zwitter distances as functions of J-K colour.}
\label{VaJmK}
\end{figure}

To obtain subsamples with different kinematics, the stars are binned according
to their J-K colours. These subsamples show a clear systematic trend with colour in the mean tangential velocity $\Delta V$ and in the radial velocity dispersion $\sigma_{R}$ (see Figure \ref{VaJmK}).
We find a larger velocity dispersion with decreasing metallicity, as expected. But in contrast to the general expectation, the corresponding asymmetric drift is decreasing with decreasing metallicity, meaning faster rotation of lower metallicity populations. This inverted trend is more pronounced in the bluer colour bins with a younger mean age of the subpopulations. A similar trend was already observed in the thin disc sample of G dwarfs from the Sloan Extension for Galactic Understanding and Exploration (SEGUE) \citep{lee11b, liu12} and for the younger population in the Geneva-Copenhagen Survey \citep{loebman11}.

We do not attempt to separate thin and thick disc stars, but due to the vertical limitation $|z|\leq 500$\,pc the thin disc is expected to dominate. Instead, we split the samples in the colour bins further into three metallicity bins, -0.5$<$[M/H]$<$-0.2, -0.2$<$[M/H]$<$0, and 0$<$[M/H]$<$0.2.
Even though the absolute calibration of the RAVE metallicity is not completely settled \citep{boeche2011},
the metallicity [M/H] from the RAVE pipeline can be used as a relative indicator of the true metallicity.
The subsample properties are collected in Table \ref{tabrave}. The total number of stars in this colour and metallicity range is $N=63\,978$, with 44\%, 47\%, and 9\% falling into the low, middle, and high metallicity bin respectively.
\begin{table*}
\begin{minipage}{140mm}
\caption{Properties of the RAVE sample.}
\label{tabrave}
\begin{tabular}{ccrrrrrrr}
\hline
J-K range & [M/H] range & $N$ & J-K & [M/H] & $R$ (kpc) & $z$ (pc) & $dR$ (pc) & $dz$ (pc) \\ \hline
          & (-0.5,-0.2) & 265 & 0.18 & -0.29 & 7.94 & -166 & 164 & 234  \\
(0.1,0.2) & (-0.2,0)    & 737 & 0.17 & -0.10 & 7.93 & -120 & 160 & 231  \\
          & (0,0.2)     & 379 & 0.16 & 0.06 & 7.92 & -113 & 181 & 236  \\  \\
          & (-0.5,-0.2) & 6105 & 0.27 & -0.31 & 7.94 & -146 & 139 & 226  \\
(0.2,0.3) & (-0.2,0)    & 9808 & 0.26 & -0.11 & 7.93 & -133 & 149 & 220  \\
          & (0,0.2)     & 3324 & 0.27 & 0.05 & 7.92 & -141 & 160 & 233  \\  \\
          & (-0.5,-0.2) & 6393 & 0.34 & -0.30 & 7.92 & -120 & 136 & 211  \\
(0.3,0.4) & (-0.2,0)    & 13917 & 0.34 & -0.10 & 7.93 & -122 & 136 & 210  \\
          & (0,0.2)     & 9787 & 0.35 & 0.07 & 7.93 & -133 & 134 & 212  \\  \\
          & (-0.5,-0.2) & 955 & 0.44 & -0.30 & 7.94 & -71 & 95 & 140  \\
(0.4,0.5) & (-0.2,0)    & 3260 & 0.44 & -0.09 & 7.94 & -82 & 104 & 158  \\
          & (0,0.2)     & 4155 & 0.44 & 0.08 & 7.94 & -97 & 110 & 173  \\  \\
          & (-0.5,-0.2) & 1331 & 0.66 & -0.30 & 7.98 & -36 & 42 & 67  \\
(0.5,0.9) & (-0.2,0)    & 3049 & 0.64 & -0.09 & 7.97 & -41 & 51 & 80  \\
          & (0,0.2)     & 2593 & 0.62 & 0.07 & 7.97 & -47 & 59 & 93  \\  \\
\end{tabular}

\medskip
Columns 1 and 2 give the colour and metallicity bins, column 3 the number of stars in the bins, columns 4 to 7 the mean values in colour, metallicity, Galactocentric distance, and vertical position respectively. Columns 8 and 9 are the mean radial and vertical distance to the Sun.
\end{minipage}
\end{table*}

From Table \ref{tabrave} we see that different bins probe slightly different volumes, with bluer bins (which correspond to brighter stars) extending farther both in radial and vertical directions.
Due to the asymmetry of the RAVE sample, the mean radius $R$ differs from $R_0=8$\,kpc and the mean height $z$ differs from 0, with the difference also being larger for bluer bins. These small variations and offset have no significant impact on the derivation of the kinematic properties at the solar position.
It is important to mention that the volume occupied by a subsample does not strongly depend on its metallicity
and that the metal-poor stars are on average about 30\% farther away from the Sun than the metal-rich stars only for the reddest bin.

\begin{figure}
  \includegraphics[height=85mm,angle=270]{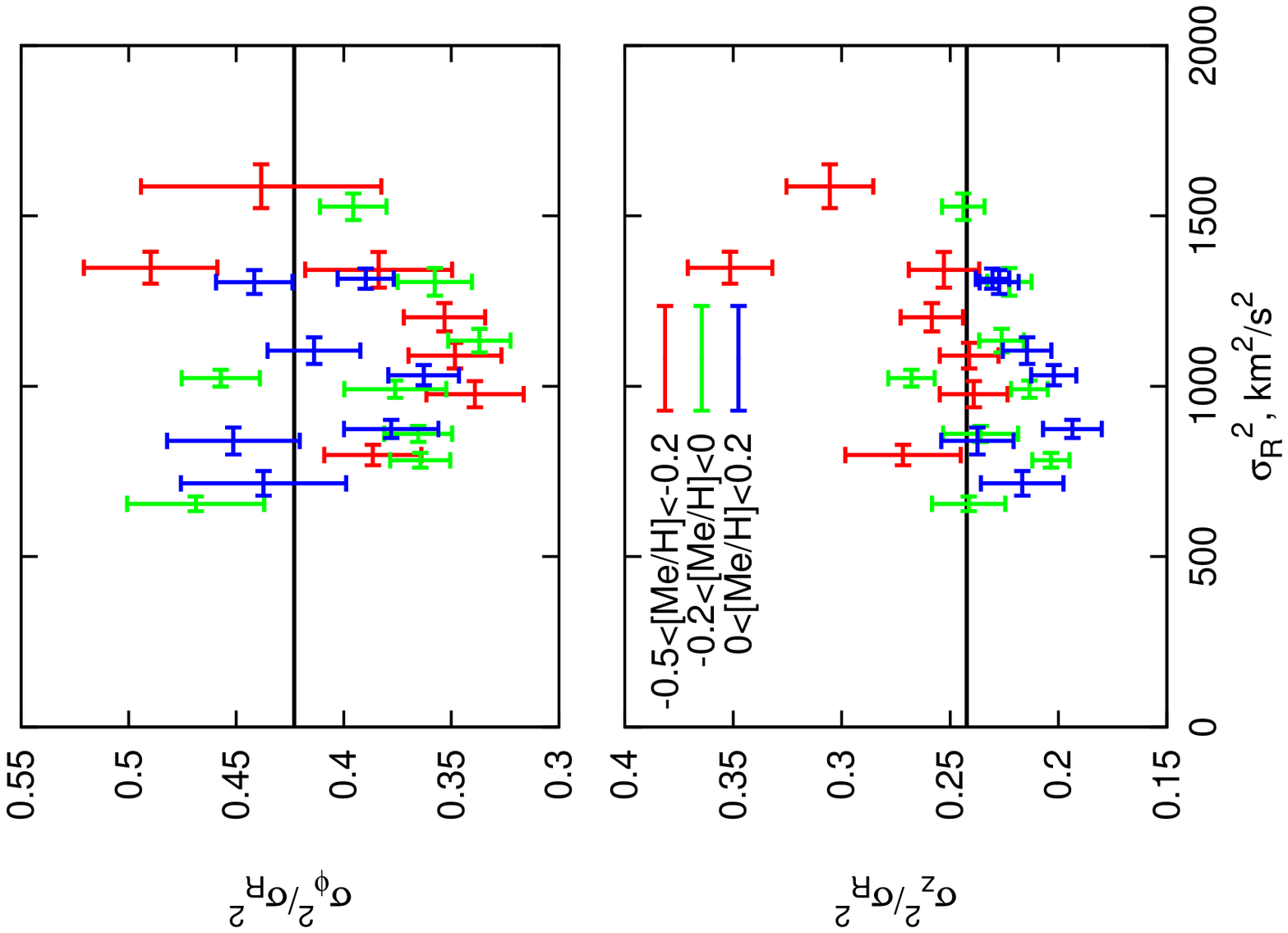}   
  \caption{Properties of the velocity ellipsoid from the RAVE data.
The squared axis ratios of the velocity
ellipsoid
$\sigma_{\phi}^2/\sigma_\mathrm{R}^2$ and $\sigma_\mathrm{z}^2/\sigma_\mathrm{R}^2$ as a function of
J-K are plotted. The mean values are marked by horizontal lines.}
\label{ellipsoid}
\end{figure}

To take full advantage of the stellar parameter estimation in RAVE, we
 measure the shape of the velocity ellipsoid.
In Figure \ref{ellipsoid} the upper panel shows the squared ratio of the velocity dispersions
in the rotational and radial directions, $\sigma_\mathrm{\phi}^2/\sigma_\mathrm{R}^2$.
There is a trend with velocity dispersion (which is discussed more in Sect. \ref{sec-discussion})
but no significant differences for different metallicities.
In the epicyclic approximation, the ratio is connected to the local rotation curve by
$\sigma_\mathrm{\phi}^2/\sigma_\mathrm{R}^2= \kappa^2/4\Omega^2\sim 0.46$
for standard values \citep{binney},
where $\kappa$ is the epicyclic frequency in the solar neighbourhood and $\Omega$ is the orbital frequency.
The observed deviations may be due to spiral structure of the Galactic disc at the low-velocity-dispersion end
and to the non-harmonic motion with respect to the guiding centre of stars with larger eccentricity
at the high-velocity-dispersion end.
In the lower panel of Figure \ref{ellipsoid} the ratio $\sigma_\mathrm{z}^2/\sigma_\mathrm{R}^2$ is presented.
We can see that the ratio is bigger for bigger velocity dispersions and for
lower metallicities. In both panels the mean values, which are used in the standard analysis in Section \ref{sec-lin}, are shown as horizontal lines.

The radial and vertical components of the LSR from the RAVE data are $U_\mathrm{\sun}=8.74\pm 0.13$ km s$^{-1}$ and $W_\mathrm{\rm
\sun}=7.57\pm 0.07$ km s$^{-1}$.
They are in reasonable agreement with $U_\mathrm{\sun}=9.96\pm 0.33$ km s$^{-1}$ and
$W_\mathrm{\sun}=7.07\pm 0.34$ km s$^{-1}$ from \citet{aumer09}.
The discrepancy of order of $1$ km s$^{-1}$ does not make a big difference in computations of velocity dispersions
as it is only added to the velocity dispersion quadratically.

\subsection{Other samples}

We used four other independent kinematically unbiased samples of dwarfs for comparison and to check the consistency of the RAVE sample with older determinations of the asymmetric drift.

1. A large independent homogeneous sample consists of F- and G-dwarfs from the Sloan Extension for Galactic Understanding and Exploration (SEGUE) \citep{segue} of the Sloan Digital Sky Survey (SDSS).
Stellar parameters, including $v_r$, metallicities [Fe/H] and alpha-abundances [$\alpha$/Fe] are computed by \citet{lee11a}.
We start with the G-dwarf sample used in \citet{lee11b}, where proper motions $\mu$ and distances $r$ were added.
For our analysis we only selected stars with a signal-to-noise ratio $\mathrm{S/N}>30$ and $\mathrm{log}\,g>4.2$ in the local volume described above.
For calculations of the propagation of errors, uncertainties of $\Delta r=0.3r$
and $\Delta v_r=4$ km s$^{-1}$ are assumed. With these criteria we got a total of $N=1190$ stars.
The majority of stars in the sample belong to a narrow colour range $0.48<g-r<0.55$,
which makes studying kinematics as a function of colour virtually impossible.

2. The sample of SEGUE M-dwarfs is taken from \citet{west11}.
It includes SDSS photometry, $v_r$, $\mu$, $T$, $\mathrm{log}g$, and photometric distances $r$.
Here only stars with distances $r<700$\,pc were selected to avoid possible velocity
biases of more distant stars \citep{bochanski11}.
We selected stars with errors $\Delta\mu<10$ mas yr$^{-1}$. Errors $\Delta
r=0.3r$ and $\Delta v_r=4$ km s$^{-1}$ were assumed. The resulting number of stars is $N=30814$.

3. The Hipparcos sample is restricted to completeness limits in $V$ magnitude bins
and supplemented by the Catalogue of Nearby Stars (CNS4)
to have a better representation of the faint end of the main sequence, as discussed in \citet{just10}, with a total of  $N=2176$ stars.
The stars have Johnson B and V photometry, $v_r$, $\mu$, and parallaxes.
The sample is binned according to absolute magnitude in $V$.

4. The last data set is a sample of McCormick K and M dwarfs
with stellar ages determined by atmosphere activities \citep{vyssotsky63}.
It contains 516 stars with reliable distances and space velocity components.
The sample is binned in stellar age.

\section{Jeans analysis}

The asymmetric drift is governed by the Jeans equation \citep{binney},
\begin{equation}
v_\mathrm{c}^2-\overline{v_{\phi}}^2=\sigma_{\phi}^2-\sigma_\mathrm{R}^2
-\sigma_\mathrm{R}^2 R \frac{
\partial\ln(\nu \sigma_\mathrm{R}^2)}{\partial R}
-R\frac{\partial(\overline{v_\mathrm{R} v_\mathrm{z}})}{\partial z},
\label{jeans}
\end{equation}
with tracer density $\nu$ and covariance $\overline{v_\mathrm{R} v_\mathrm{z}}$.
Roughly speaking, it expresses dynamical equilibrium in an axisymmetric
system within a volume element in a cylindrical coordinate system.
The left-hand side represents the difference of the gravitational force in the Galactic
potential and  the centrifugal force, while
the terms on the right-hand side represent dynamical pressure and shear forces acting on
the surfaces of the volume.
There are two crucial  assumptions for the validity of Eq.
(\ref{jeans}),
namely axisymmetry of the system and dynamical
equilibrium of the stellar population under consideration.
The former assumption can be broken by a spiral density wave, while the latter
can be violated for young populations,
whose mean age is smaller than the epicyclic period.

In Eq. (\ref{jeans}) the radial gradient term can be parameterised by the local radial scalelength $R_\mathrm{E}$ of $(\nu \sigma_\mathrm{R}^2)$ via $R_\mathrm{E}^{-1}=\partial\ln(\nu \sigma_\mathrm{R}^2)/\partial R$. It is a composition of the radial scalelength $R_\mathrm{\nu}$ of the tracer density $\nu$ and $R_\mathrm{\sigma}$ of the radial velocity dispersion $\sigma_\mathrm{R}^2$ related by $R_\mathrm{E}^{-1}=R_\mathrm{\nu}^{-1}+R_\mathrm{\sigma}^{-1}$.

The vertical gradient of the covariance $\overline{v_\mathrm{R} v_\mathrm{z}}$ in Eq. (\ref{jeans}) measures the orientation of the principal axes of the velocity ellipsoid above and below the Galactic plane. We use the parametrisation $R\partial(\overline{v_\mathrm{R} v_\mathrm{z}})/\partial z=\eta \left(\sigma_\mathrm{R}^2-\sigma_\mathrm{z}^2 \right)$; $\eta=0$ corresponds to a horizontal orientation of the principal axes and $\eta=1$ to a spherical orientation.

Finally, we replace $v_\mathrm{c}$ and $\overline{v_{\phi}}$ by  $v_\mathrm{\sun}$, $V_\mathrm{\sun}$, and $\Delta V$ and evaluate Eq. (\ref{jeans}) at the solar position $R=R_0$. Re-arranging the terms and dividing by $2v_\mathrm{\sun}$ we find
\begin{eqnarray}
V_\mathrm{a}&=&v_\mathrm{c}-\overline{v}_\mathrm{\phi}=\Delta V-V_\mathrm{\sun} \label{nonlinear}\\
&=& \frac{\Delta V^2 - V_\mathrm{\sun}^2}{2v_\mathrm{\sun}} + \frac{\sigma_\mathrm{R}^2}{2v_\mathrm{\sun}}
\left(\frac{R_0}{R_\mathrm{E}}-1 - \eta\left[1-\frac{\sigma_\mathrm{z}^2}{\sigma_\mathrm{R}^2}\right] + \frac{\sigma_\mathrm{\phi}^2}{\sigma_\mathrm{R}^2}\right)
\nonumber
\end{eqnarray}
This is the non-linear equation for the asymmetric drift $V_\mathrm{a}$ as function of $\sigma_\mathrm{R}^2$ for a set of stellar subpopulations. It connects the measured mean tangential velocity $-\Delta V=\overline{v}_\mathrm{\phi} - v_\mathrm{\sun}$ of a subpopulation with respect to the Sun and the peculiar motion of the Sun $V_\mathrm{\sun}$.
There are two types of non-linearity on the right-hand side of equation \ref{nonlinear}. The two quadratic terms $\Delta V^2$ and $V_\mathrm{\sun}^2$ yield a small correction to the asymmetric drift with increasing significance of the first one with increasing velocity dispersion (e.g. for the thick disc). This correction can easily be taken into account.
The second non-linearity is more crucial and occurs from a possible variation of the radial scalelength and the shape and orientation of the velocity ellipsoid for the different subpopulations introducing an additional dependence of the last bracket in  equation
\ref{nonlinear} on $\sigma_\mathrm{R}$.

Since the thickness of a stellar tracer population depends on the total surface density and the vertical velocity dispersion, a radially independent constant thickness requires a constant shape of the velocity dispersion ellipsoid to find $R_\mathrm{\sigma}=R_\mathrm{d}$, the scalelength of the total surface density. In the simplest case, where the radial scalelength of the tracer population $R_\mathrm{\nu}$ is the same, i.e. $R_\mathrm{d}=R_\mathrm{\nu}=R_\mathrm{\sigma}$, we get $R_\mathrm{\nu}=2R_\mathrm{E}$ in the asymmetric drift equation \ref{nonlinear}.

The impact of the orientation of the velocity ellipsoid via $\eta$ in the Jeans equation is twofold. Since $\sigma_\mathrm{z}<\sigma_\mathrm{R}$, a spherical orientation ($\eta=1$) results in a smaller asymmetric drift compared to a horizontal orientation with $\eta=0$ . On the other hand, a stellar population with a measured asymmetric drift $V_\mathrm{a}$ requires a larger radial scalelength $R_\mathrm{E}$ for $\eta=1$ to fulfill the Jeans equation.
For definiteness we adopt $\eta=1$ \citep[supported observationally and theoretically by][]{siebert08,binney10} in the plots and interpretation of data, if necessary.

\subsection{The linear Str\"omberg relation}

In the standard application the quadratic terms $\Delta V^2$ and $V_\mathrm{\sun}^2$ in Eq. (\ref{nonlinear}) are neglected and we find Str\"{o}mberg's equation
\begin{eqnarray}
\Delta V&=&V_\mathrm{\sun}+\frac{\sigma_\mathrm{R}^2}{k}
\nonumber\\
k&=&2v_\mathrm{\sun}
\left(\frac{R_0}{R_\mathrm{E}}-1 - \eta\left[1-\frac{\sigma_\mathrm{z}^2}{\sigma_\mathrm{R}^2}\right] + \frac{\sigma_\mathrm{\phi}^2}{\sigma_\mathrm{R}^2}\right)^{-1}.
\label{stromberg}
\end{eqnarray}
The  inverse slope $k$ depends on the radial scalelength $R_\mathrm{E}$ and shape and orientation of the velocity dispersion ellipsoid of the subpopulations with density $\nu$.
If we assume that the shape and orientation of the velocity ellipsoids are the same, i.e.
$\sigma_\mathrm{R}^2 \propto \sigma_\mathrm{\phi}^2 \propto \sigma_\mathrm{z}^2$ and same $\eta$, and that the radial scalelength is the same for all subpopulations, then $k$ is the same for all subpopulations and thus independent of $\sigma_\mathrm{R}$. With these assumptions we end up with the classical linear Str\"omberg relation, which we discuss in more detail in section \ref{sec-lin}.

\subsection{A new Str\"omberg relation}

Since we have measurements of the shape of the velocity ellipsoid for each
subsample, it is useful to separate observables and unknowns in the non-linear asymmetric drift
equation (\ref{nonlinear}) by rewriting it as
\begin{equation}
V'=V_\mathrm{\sun}-\frac{V_\mathrm{\sun}^2}{2v_\mathrm{\sun}}+\frac{\sigma_\mathrm{R}^2}{k'}
\label{new-stromberg}
\end{equation}
with
\begin{eqnarray}
V'&\equiv&\Delta V+\frac{\sigma_\mathrm{R}^2 + \eta(\sigma_\mathrm{R}^2-\sigma_\mathrm{z}^2) -\sigma_\mathrm{\phi}^2-\Delta V^2}{2v_\mathrm{\sun}},
\label{Vprime}
\end{eqnarray}
\begin{eqnarray}
k'&=&v_\mathrm{\sun}\left(\frac{2R_\mathrm{E}}{R_0}\right).
\label{Kprime}
\end{eqnarray}
The new quantity $V'$ contains corrections arising from the shape and orientation of the velocity ellipsoid and the quadratic term $\Delta V^2$. The quadratic $V_\mathrm{\sun}^2$ term on the right-hand side of Eqn. (\ref{new-stromberg}) decreases the zero point $V'(\sigma_\mathrm{R}=0)$ with respect to the value of $V_\mathrm{\sun}$ by 1--2\%.
The new parameter $k'$ depends only on the radial scalelength $R_\mathrm{E}$ of the stellar subpopulations.
In the new form of the Str\"omberg relation we need to assume only equal radial scalelengths
for a linear fit to the data to determine the peculiar motion of the sun $V_\mathrm{\sun}$ by the zero point and $R_\mathrm{E}$ via the inverse slope $k'$. In general,
the scalelength $R_\mathrm{E}$ could be also a function of $\sigma_\mathrm{R}$, thus
implying a dependence of $k'$ on $\sigma_\mathrm{R}$ in Eq. (\ref{new-stromberg}). It is discussed in detail in section \ref{sec-new}.

\section{Results}\label{sec-results}

We discuss first the application of the linear Str\"omberg relation on the RAVE data in comparison with the other data sets. In a second step we repeat the analysis with the RAVE data split into the metallicity bins and then apply the new Str\"omberg relation. In the third step we investigate in a more general frame the determination of the LSR and the radial scalelengths. Finally, we discuss a very simple toy model, which can reproduce our findings.

\subsection{Standard analysis}\label{sec-lin}

\begin{figure}
  \includegraphics[height=85mm,angle=270]{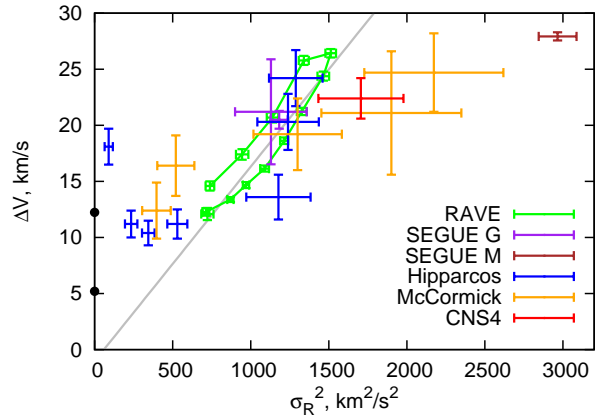}
  \caption{Asymmetric drift for different data sets.
  The two black circles on the $\Delta V$ axis correspond to the different LSRs with $V_\mathrm{\sun}=5.25$ km s$^{-1}$ from \citet{aumer09}
  and $V_\mathrm{\sun}=12.24$ km s$^{-1}$ from \citet{schoenrich10} respectively.
  The grey line gives the best fit to the data points for RAVE dwarfs.
  It corresponds to the LSR $V_\mathrm{\sun}=-1.04$ km s$^{-1}$ and the scalelength of the disc $R_\mathrm{d}=1.65$\,kpc.}
\label{fig1}
\end{figure}

In Figure \ref{fig1} we see that the linear Str\"omberg relation (Eq. \ref{stromberg}) with constant slope $k^{-1}$ is poorly applicable to the RAVE data:
the data points are not following the same straight line.
The formal best fit to the RAVE data (grey line in Figure \ref{fig1}) gives the LSR $V_\mathrm{\rm
\sun}=-1.0\pm 2.1$ km s$^{-1}$,
which is not consistent with $V_\mathrm{\sun}=5.25\pm 0.54$ km s$^{-1}$
obtained by \citet{aumer09} by a similar linear fit to Hipparcos data. The corresponding
slope $k=58\mathrm{km\,s}^{-1}$ is bigger than the
classical value.
An application of Eq. (\ref{stromberg}) with the mean ratios of the squared
velocity dispersions ($\sigma_\mathrm{\phi}^2/\sigma_\mathrm{R}^2=0.42$
and $\sigma_\mathrm{z}^2/\sigma_\mathrm{R}^2=0.24$, see Figure \ref{ellipsoid}) results in a
short radial scalelength of $R_\mathrm{d}=2R_\mathrm{E}=1.65 \pm 0.16$\,kpc.

The SEGUE G dwarfs allow us to get only one significant point in the plot, and this point is consistent with the trend obtained from RAVE,
while SEGUE M dwarfs seem to be off the trend. The M dwarf sample may suffer from biases in the distance determination.
The local stars from the Hipparcos, CNS4, and McCormick samples
are also generally consistent with the best-fitting line for RAVE,
except for the two dynamically coldest bins. This feature, which was already observed by \citet{dehnen98}, could be explained by the fact that the young stars have not yet
reached dynamical equilibrium.

Overall, the SEGUE G dwarfs and the Hipparcos data support the slope $k^{-1}$ determined by the RAVE data but with much larger scatter.
The McCormick stars, the SEGUE M dwarfs, and the CNS4 data suggest a much smaller slope and larger LSR value, which would be inconsistent with the RAVE sample but support the large LSR value claimed by \citet{schoenrich10}.
The increase of the observed $V_\mathrm{a}$ (or equivalently $\Delta V$) in Figure \ref{fig1} for the smallest velocity dispersions is inconsistent even with the model by \citet{schoenrich10}, suggesting non-equilibrium of the young subpopulation.

\subsection{Metallicity dependence}\label{sec-met}

Iin the Hipparcos sample a non-linear trend of the asymmetric drift with increasing velocity dispersion has already been observed. This is a sign that the radial scalelength is different for different subpopulations. Numerical models of Milky-Way-like galaxies also predict a systematic variation of the asymmetric drift with age and/or metallicity
\citep[e.g.][]{schoenrich10,loebman11}.
\citet{schoenrich09} have shown that radial mixing leads to a slight increase of the radial scalelength with increasing age.
In \citet{lee11b} it was shown for the SEGUE G dwarf sample that the asymmetric drift in the thin disc decreases with decreasing metallicity in contrast to the naive expectation of local evolution models.
\citet{bovy12} also used the full SEGUE G dwarf sample (dominated by stars with $|z|>500$\,pc) to derive radial scalelengths of mono-abundance subpopulations. They found a significantly smaller radial scalelength for thick disc stars compared to thin disc stars, with a hint of decreasing scalelength with increasing metallicity inside the thin disc.

\begin{figure}
  \includegraphics[height=85mm,angle=270]{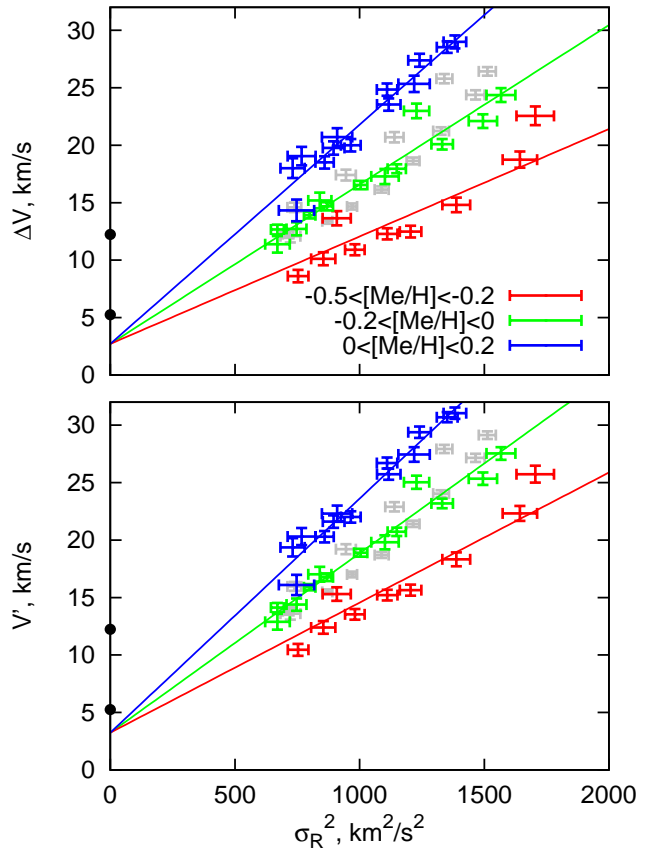}   
  \caption{The asymmetric drift for the RAVE dwarfs separated into three metallicity bins: $-0.5<$[M/H]$<-0.2$, $-0.2<$[M/H]$<$0, and 0$<$[M/H]$<$0.2.
  The two black circles on the $y$-axis correspond to the LSR from
\citet{aumer09} and from \citet{schoenrich10}.
  The full lines show the best joint linear fit.
  Top: Using Eq. (\ref{stromberg}). Bottom: Using Eq. (\ref{new-stromberg}).
The RAVE data without metallicity split of Figure \ref{fig1} are replotted with grey points.}
\label{metallicity}
\end{figure}

The RAVE sample in Figure \ref{fig1} also shows a systematic non-linear trend, which may be due to a varying mixture of different populations with different scalelengths.
Binning stars of the RAVE sample in metallicities allow us to see more
interesting features in the behaviour of the asymmetric drift.
Figure \ref{VaJmK} shows that there is a systematic trend in both the velocity dispersion and the asymmetric drift with metallicity, which is in part due to the bluer colour of more metal-poor stars.
In the top panel of Figure \ref{metallicity} we plot the mean rotational velocity
$\Delta V$ versus its squared radial velocity dispersion $\sigma_\mathrm{R}^2$
for the three different metallicity bins, -0.5$<$[M/H]$<$-0.2, -0.2$<$[M/H]$<$0, and 0$<$[M/H]$<$0.2.
We see that stars at different metallicities demonstrate different asymmetric drifts,
with more metal-poor stars having smaller asymmetric drifts and thus larger rotational velocities. For comparison the RAVE data from Figure \ref{fig1} are replotted in grey to demonstrate that the non-linearity is partly resolved by the separation into metallicity bins.

The common LSR $V_\mathrm{\sun}$ and the three inverse slopes $k$ for each metallicity bin are the free parameters for a joint linear fit of the asymmetric drift equation  \ref{stromberg}.
In the top panel of Figure \ref{metallicity} the best joint linear fit is shown.
We find for the LSR $V_\mathrm{\rm \sun}=2.52\pm 0.80$ km s$^{-1}$,
which is consistent with the estimate from Figure \ref{fig1}. The radial scalelengths of the three metallicity components can be estimated from the inverse slopes $k$ by
inserting the mean ratios of the squared
velocity dispersions ($\sigma_\mathrm{\phi}^2/\sigma_\mathrm{R}^2=0.41, 0.40, 0.42$
and $\sigma_\mathrm{z}^2/\sigma_\mathrm{R}^2=0.28, 0.23, 0.22$, for the low, intermediate, and high metallicity sample respectively).
The radial scalelengths are $2.73 \pm 0.17$, $1.97 \pm 0.10$, and $1.50 \pm 0.05$\,kpc with increasing metallicity, assuming $R_\mathrm{\nu}=R_\mathrm{\sigma}$. The decreasing radial scalelength with increasing metallicity corresponds to a negative radial metallicity gradient because the fraction of metal-poor stars increases with increasing radius.

Now we relax the assumption of similar velocity dispersion ellipsoids of the different colour-metallicity bins and apply the new Str\"omberg relation  derived in Eq. (\ref{new-stromberg}). In the bottom panel of Figure \ref{metallicity}, $V'$ as function of $\sigma_\mathrm{R}^2$ is plotted. All data points are shifted up by a few km\,s$^{-1}$, but the general picture does not change. The inverse slopes $k'$ of the joint linear regression are now a direct measure of the radial scalelengths of the stellar populations in the different metallicity bins.
We find for the LSR $V_\mathrm{\sun}=3.06\pm 0.68$ km s$^{-1}$ slightly larger than the previous value.
The radial scalelengths of the disc are $2.91 \pm 0.16$, $2.11 \pm 0.09$, and $1.61 \pm 0.05$\,kpc with increasing metallicity. The systematically
larger radial scalelengths in the new analysis (bottom panel of Figure \ref{metallicity}) are mostly due to the shift of the LSR. The similarity of the classical and new analysis demonstrates the small impact of the velocity ellipsoid compared to the radial scalelength term in the asymmetric drift equation.
Adopting a horizontal orientation of the velocity dispersion ellipsoid $\eta=0$ in Eq.~(\ref{stromberg}) yields slightly larger scalelengths of $3.11\pm 0.23$, $2.18\pm 0.12$, and $1.62\pm 0.23$\,kpc respectively.

We can use these radial scalelengths to estimate the metallicity gradient in the disc. We assume the disc to consist of three populations, whose densities are described by exponentials with the corresponding scalelengths. Their metallicities are assumed to be $-0.35$, $-0.1$, and $0.1$, which are median metallicities of the adopted bins. Relative weights of the populations at the solar radius are taken proportional to the total number of stars in the corresponding bins (see Table \ref{tabrave}). We get a shallow radial metallicity gradient of $-0.016\pm 0.002$ dex\,kpc$^{-1}$. To reproduce the observed metallicity gradient of $-0.051\pm 0.005$ dex\,kpc$^{-1}$ \citep{coskunoglu12}, a much larger range of radial scalelengths or metallicities is required.

\subsection{The LSR and radial scalelengths}\label{sec-new}

In the previous analysis, we still adopted the same radial scalelength in each metallicity bin independent
of the colour $J-K$ and thus of the velocity dispersion. If we also relax this assumption, as suggested in the literature mentioned in Section \ref{sec-met}, then it is no longer possible to determine the LSR (i.e.$V_\mathrm{\sun}$) by a linear extrapolation to $\sigma_\mathrm{R}^2=0$. For any extrapolation we would need a prediction of the dependence  $V'(\sigma_\mathrm{R}^2)$, e.g. from a model.

\begin{figure}
  \includegraphics[height=85mm,angle=270]{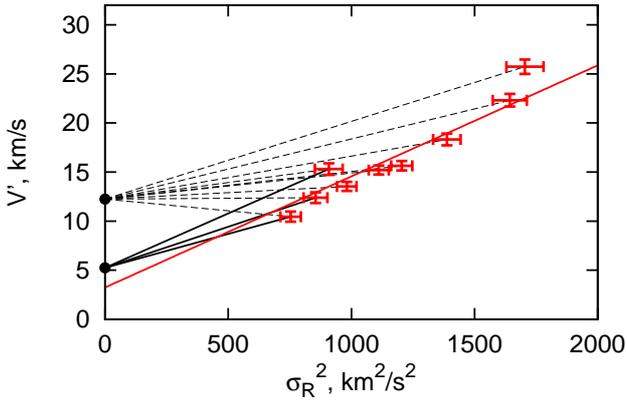}   
  \caption{Same data for the low-metallicity bin as in the bottom panel of Figure \ref{metallicity}.
  The two black circles on the $y$-axis correspond to the LSR from
\citet{aumer09} and from \citet{schoenrich10}.
  The full and dashed lines indicate for some subsamples the individual slopes and their dependence on the adopted LSR values, which are proportional to the inverse radial scalelengths.}
\label{metallicity2}
\end{figure}

Instead we may adopt a value for $V_\mathrm{\sun}$ and derive individual radial scalelengths $R_\mathrm{E}(\nu)$ for each data point by determining the parameter $k'(\nu)$. The parameter $k'(\nu)$ corresponds to the inverse slope of the line connecting the data point with the zero point of equation \ref{new-stromberg}.
This is demonstrated in Figure \ref{metallicity2} for a few data points and the LSR of \citet[][full black lines]{aumer09} compared to that of  \citet[][dashed black lines]{schoenrich10}. The connecting lines are no longer linear fits to data but a visualization of $k'(\nu)$ from the application of equation \ref{new-stromberg} to each data point.
Lifting the LSR value results in larger $k'(\nu)$ (smaller slopes) for all subsamples, leading to larger radial scalelengths from equation \ref{Kprime}.

\begin{figure}
  \includegraphics[height=85mm,angle=270]{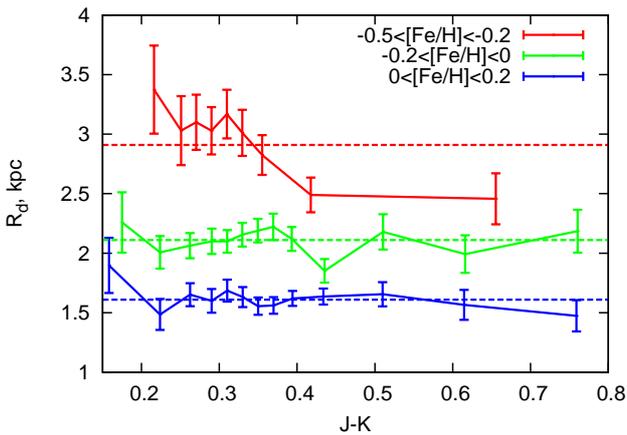}   
  \caption{Radial scalelengths corresponding to the best fit in different bins, calculated for the LSR $V_\mathrm{\sun}=3.06$ km s$^{-1}$ . Horizontal dashed lines represent the radial scalelengths used in the best fit in the lower panel of Figure \ref{metallicity}.}
\label{Rd_vs_JmK}
\end{figure}

Figure \ref{Rd_vs_JmK} shows the variation of the radial scalelength as a function of colour $J-K$ for the 
different metallicity bins, adopting the best-fit value for the solar motion $V_\mathrm{\sun}=3.06$ km s$^{-1}$.
For the higher metallicity bins, the data are consistent with a constant radial 
scalelength for all stars along the main sequence. In the low-metallicity bin a 
significant decline of $R_\mathrm{d}$ with the mean age of the stars is obvious in the sense of 
larger radial scalelength for the young metal-poor subpopulation.

\begin{figure*}
  \includegraphics[height=170mm,angle=270]{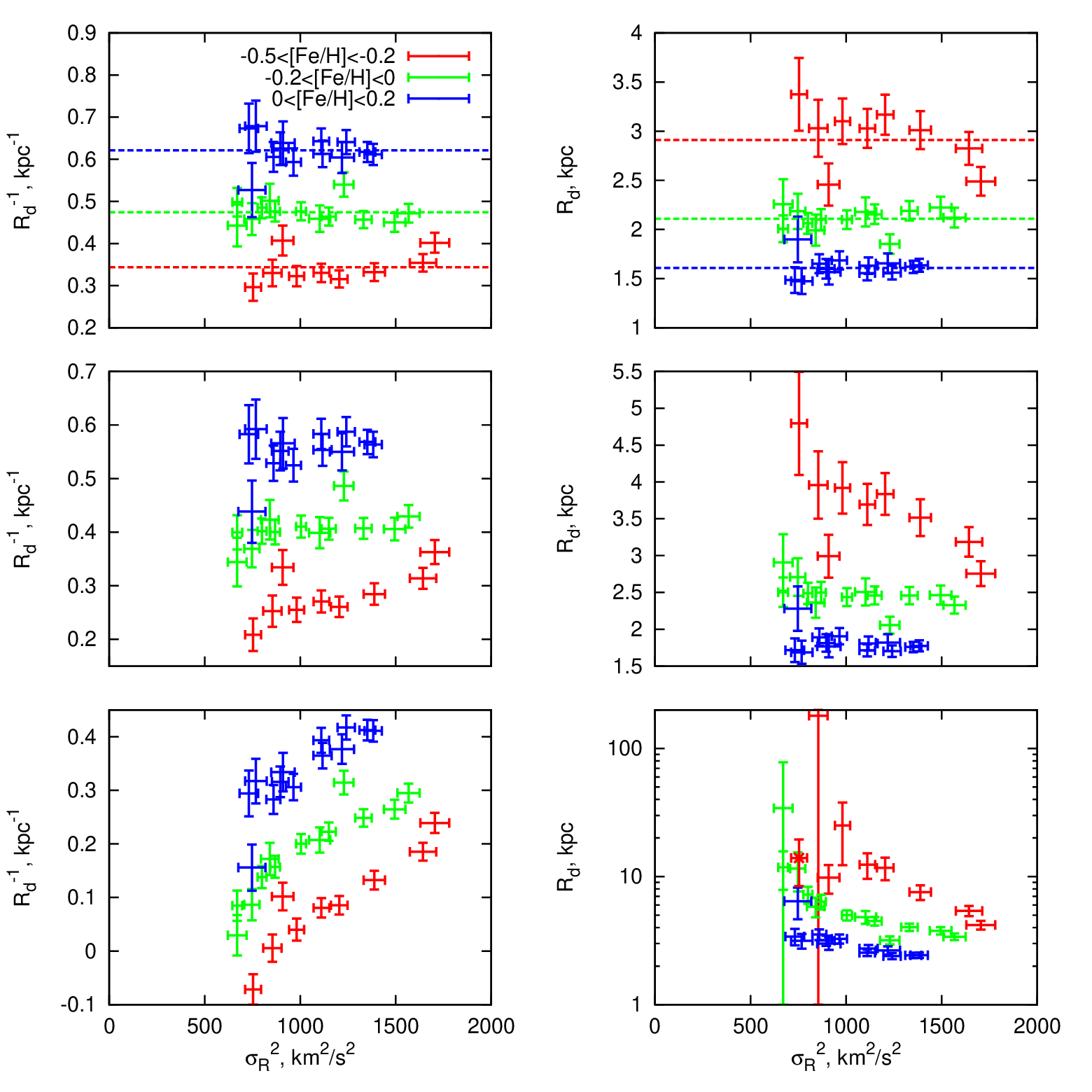}
  \caption{Left panels: Inverse radial scalelengths for all subpopulations adopting the LSR $V_\mathrm{\sun}=3.06$ km s$^{-1}$ (best fit in the lower panel of Figure \ref{metallicity}, top panel),  $5.25$ km s$^{-1}$ (\citet{aumer09}, middle panel), and $12.24$ km s$^{-1}$ (\citet{schoenrich10}, lower panel) with colour-coded metallicity.
Right panels: Radial scalelengths for the same data. For the Sch\"onrich value of the LSR (lower panel), the absolute values $|R_\mathrm{\nu}|$ are plotted in logarithmic scale with negative values marked as crosses.}
\label{Rd-1}
\end{figure*}

The left-hand panels of Figure \ref{Rd-1} show the inverse radial scalelengths for all RAVE subpopulations adopting the best fit LSR $V_\mathrm{\sun}=3.06$ km s$^{-1}$ (top panel), and the LSR of \citet{aumer09} (middle panel) and \citet{schoenrich10} (lower panel), with colour-coded metallicity. The right-hand panels show the corresponding  radial scalelengths $R_\mathrm{\nu}=R_\mathrm{\sigma}$. Since the LSR of \citet{schoenrich10} is larger than some $V'$ values, negative values corresponding to a radially increasing density appear. More precisely, the scalelength $R_\mathrm{E}$ of $\nu\sigma_\mathrm{R}^2$ becomes negative, meaning an increasing radial energy density with increasing distance to the Galactic centre (see Equation \ref{nonlinear}). This is physically possible, for example, for metal-poor stars if the younger population born at larger radii dominates over the older stars born further in.

We observe in both cases that the radial scalelengths are systematically larger for smaller metallicity. But the trend in each metallicity bin as a function of $\sigma_\mathrm{R}^2$ depends sensitively on the adopted value for the LSR.

\subsection{A simple model}\label{sec-mod}

There is a dynamical connection between radial gradients in the disc and the asymmetric drift due to the epicyclic motion of stars on non-circular orbits. Stars with guiding radii further in show a smaller tangential velocity in the solar neighbourhood because of the vertical component of angular momentum conservation. A negative radial density gradient results in a larger fraction of stars coming from the inner part of the disc compared to the outer part. Therefore the mean tangential velocity is smaller than the local circular velocity, corresponding to a positive asymmetric drift, and the distribution in $v_\mathrm{\phi}$ is skewed. Additionally, with increasing radial velocity dispersion the average distance to the guiding radius of stars in the solar neighbourhood increases, leading to an increasing asymmetric drift with increasing $\sigma_\mathrm{R}$.  As a second effect, a negative radial gradient in $\sigma_\mathrm{R}$ further increases the asymmetric drift and the skewness.

If there is a negative metallicity gradient in the
Milky Way disc \citep[e.g. as found by][also using RAVE dwarfs]{coskunoglu12}, then
a higher fraction of metal-rich stars observed in the solar neighbourhood
is expected to possess guiding radii
smaller than $R_0$.
It means that we are observing  a larger asymmetric drift for these stars compared to more metal-poor stars at the same $\sigma_\mathrm{R}$.
In terms of Eq. (\ref{stromberg}) it means that metal-rich stars are more
centrally concentrated and have a smaller disc scalelength $R_\mathrm{\nu}$,
while metal-poor stars have a bigger scalelength $R_\mathrm{\nu}$.
Any mixing process (by the epicyclic motion, radial migration due to orbit diffusion, or resonant scattering) tends to smear out gradients and increase the local scatter.

\begin{figure}
  \includegraphics[height=85mm,angle=270]{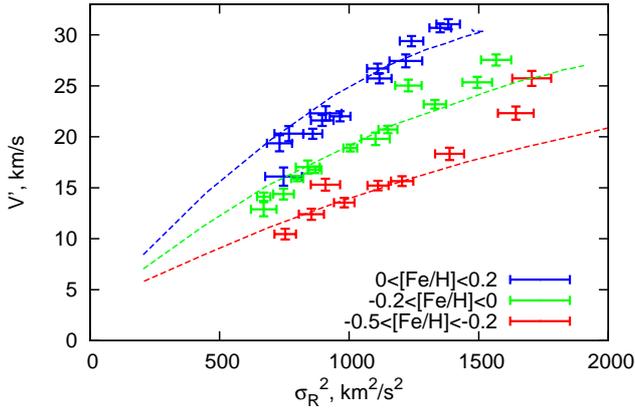}   
  \caption{Model predictions of a simple disc evolution model compared to the RAVE data shown in Figure \ref{Rd-1}.}
\label{fig-model}
\end{figure}

We demonstrate that a simple evolutionary model of the extended solar neighbourhood combining the metal enrichment and a radial metallicity gradient can reproduce a decreasing radial scalelength with increasing metallicity consistent with the observed asymmetric drift. We adopt
SFR$(R,t)\propto \exp(-R/R_\mathrm{d})$ and constant in time $t$,
the age velocity dispersion relation AVR with $\sigma_\mathrm{R}^2\propto t\exp(-R/R_\mathrm{\sigma})$,
and metal enrichment [M/H]$(R,\tau)= const +M_\mathrm{\tau}\tau+M_\mathrm{R}R \pm \Delta$[M/H] linear in age $\tau$ and in radius and allowing for a metallicity scatter. With Monte Carlo realisations for each parameter set,  we calculate the asymmetric drift and velocity dispersion for each age-metallicity bin. The result of the best-fitting parameter set with $(R_\mathrm{d},R_\mathrm{\sigma},M_\mathrm{\tau},M_\mathrm{R},\Delta\mathrm{[M/H]})$ = (1.8kpc, 1.5kpc, 0.04dex\,Gyr$^{-1}$, -0.07dex\,kpc$^{-1}$, 0.18dex) is shown in Figure \ref{fig-model}.
Even though this plot is not enough to tightly constrain all free parameters of the model,
it is educating to see how easily the observed metallicity trend can emerge.

\section{Discussion}\label{sec-discussion}

The extended, kinematically unbiased catalogue of RAVE stars provides a very good tool
to analyse stellar dynamics in the solar neighbourhood
and to study the asymmetric drift. We analysed dwarf stars selected by a colour-dependent magnitude cut.
The observed dependence of the asymmetric drift velocity $V_\mathrm{a}$ on the squared radial velocity dispersion $\sigma_\mathrm{R}^2$
is substantially non-linear, and the linear Str\"omberg relation fails to give a reasonable approximation of the data.
A somewhat similar analysis of the RAVE data was performed by \citet{coskunoglu11}.
The authors used a kinematically selected sample of stars with photometric distances
to determine the velocity of the Sun with respect to the neighbouring stars.
The mean velocity of the Sun of about 13 km s$^{-1}$ with respect to the local stars determined by \citet{coskunoglu11}
is consistent with the mean $\Delta V$ for the RAVE stars in Figure \ref{fig1}.
However, \citet{coskunoglu11} could not decompose this velocity into the peculiar velocity of the Sun with respect to the LSR
and the asymmetric drift, which is the velocity of the LSR with respect to the mean velocity of stars in their sample. Therefore they did not derive $V_\mathrm{\sun}$ but only  $(V_\mathrm{\sun}+V_\mathrm{a})$, in contrast to their suggestion.

When splitting the RAVE sample into three metallicity bins, the non-linearity of the asymmetric drift is reduced in each metallicity bin and a joint best linear fit confirms the low peculiar velocity of the Sun $V_\mathrm{\sun}=2.52\pm 0.80$ km s$^{-1}$. The slopes of the asymmetric drift yield radial scalelengths of $2.73 \pm 0.17$, $1.97 \pm 0.10$, and $1.50 \pm 0.05$\,kpc with increasing metallicity using the average values for the velocity dispersion shape.

For modern large samples like RAVE and SDSS/SEGUE, space velocities are available by combining the survey data with distance estimates and proper motion catalogues. Therefore the velocity dispersion ellipsoid is available for each stellar subsample, and we propose to rearrange the Jeans equation in such a way that all measured contributions are combined on the left-hand side to $V'$. This leads to an improved asymmetric drift equation (Eq.~\ref{new-stromberg}). In this new Str\"omberg equation, the only unknown is the radial  scalelength $R_\mathrm{E}$ of $\nu\sigma_\mathrm{R}^2$, which determines the slope of $V'$ as function of $\sigma_\mathrm{R}^2$. This new equation allows a cleaner investigation of the interrelation of radial scalelengths and the adopted (or determined) LSR (i.e. the peculiar motion of the Sun $V_\mathrm{\sun}$).

We applied the new Str\"omberg equation to the RAVE data split in the metallicity bins.
The best joint linear fit gives a value $V_\mathrm{\sun}=3.06\pm 0.68$ km s$^{-1}$ for the LSR.
The radial scalelengths are $2.91 \pm 0.16$, $2.11 \pm 0.09$, and $1.61 \pm 0.05$\,kpc respectively for the metallicities
[M/H]=-0.35, -0.1, and +0.1 dex in the disc. Adopting a horizontal orientation of the velocity ellipsoids above and below the midplane yield 10--20 percent larger scalelengths.
The small differences of the new and old Str\"omberg equations show that the contribution of the velocity ellipsoid terms to the asymmetric drift are less significant,
the overall trend of the asymmetric drift is dominated by the disc scalelengths and variations of it.
The radial scalelength of the disc is smaller for higher metallicities, implying a more centrally concentrated distribution of metal-rich stars.
The dependence of the asymmetric drift on metallicity can serve as a good
constraint for chemodynamical models of the Milky Way
and for the effect of radial migration on the stellar dynamics and abundance distribution in the solar
neighbourhood.

If the radial scalelengths of the subpopulations are
different for different velocity dispersions, the new Str\"omberg equation Eq. (\ref{new-stromberg}) is still
applicable,
but now the inverse slope $k'$ is no longer constant but depends on the squared velocity dispersion $\sigma_\mathrm{R}^2$ of the subpopulation.
The thus observed or theoretically predicted asymmetric drift and velocity dispersions serve us as a measure
of $k'$ on the right-hand side of Eq. (\ref{new-stromberg}), corresponding to the inverse radial scalelength $R_\mathrm{E}$ of $\nu\sigma_\mathrm{R}^2$ if the peculiar motion of the Sun $V_\mathrm{\sun}$ is known.
In addition to the overall trend of larger radial scalelengths for lower metallicities, we find within the metal-poor bin a trend of decreasing scalelength with increasing velocity dispersion. This can be a hint of an increasing contribution of thick disc stars combined with a small thick disc radial scalelength.
Alternatively, it is the contribution of a young 
metal-poor subpopulation of the thin disc with large radial scalelength (Figure \ref{Rd_vs_JmK}).
The inverted trend of faster rotation for more metal-poor stars, at least in the younger thin disc, as observed in \citet{lee11b,liu12,loebman11}, can be dynamically understood by the rule: lower metallicity $\to$ larger velocity dispersion and larger radial scalelength $\to$ smaller asymmetric drift $\to$ faster mean rotation.
The chemodynamical model by \citet{schoenrich10} probably can be
interpreted in these terms.
Each point of the non-linear dependence $V_\mathrm{a}(\sigma_\mathrm{R}^2)$ from \citet{schoenrich10} should correspond
by Eq. (\ref{new-stromberg}) via its own $k'$ to the radial scalelength $R_\mathrm{E}$.
Therefore, the dependence $V_\mathrm{a}(\sigma_\mathrm{R}^2)$ from \citet{schoenrich10} can be interpreted
as an increase of $R_\mathrm{\nu}$ and/or $R_{\sigma}$ with the velocity dispersion
$\sigma_\mathrm{R}$ of the subpopulations.
We have shown that elevating $V_\mathrm{\sun}$ to 12 km s$^{-1}$ results in significantly increased radial scalelengths, which are even negative for some low-metallicity bins. The physical interpretation of an increasing pressure $\nu\sigma_\mathrm{R}^2$ with radius is questionable. A second effect of a larger LSR value is a systematic trend of decreasing scalelength with increasing velocity dispersion. This is counterintuitive to the impact of radial migration, which should flatten radial gradients with increasing age and velocity dispersion.

Another possible explanation for the discrepancies in the determination of the LSR are non-axisymmetric features. A local spiral wave perturbation, which could influence the stellar dynamics in the solar neighbourhood, can account for an offset of $\approx 6$ km s$^{-1}$ \citep{siebert}.
It would break the axisymmetry of the gravitational potential required by Eq.
\ref{jeans},
thus making all further analysis inapplicable.
The dynamically coldest subpopulations of stars are the most susceptible to
small gravitational perturbations,
while dynamically hotter subpopulations are less affected by them.
Thus a Jeans analysis could break down for small $\sigma_\mathrm{R}^2$ while still being
a good approximation for big $\sigma_\mathrm{R}^2$. This would apply to the bluest bins of the Hipparcos sample and also to the maser measurements of star-forming regions.
There is still no precise model to correct for these effects in the solar
neighbourhood.

From the slope of the asymmetric drift dependence on the radial velocity dispersion,
we can estimate the radial scalelength $R_\mathrm{E}$ of $\nu\sigma_\mathrm{R}^2$ in the Galactic disc.
With $R_\mathrm{E}^{-1}=R_\mathrm{\nu}^{-1}+R_{\sigma}^{-1}$ and the standard assumption $R_{\sigma}=R_\mathrm{\nu}$, we get $R_\mathrm{\nu}$ ranging from 2.9 kpc to 1.6 kpc.
If $R_{\sigma}$ is significantly larger than $R_\mathrm{\nu}$, as assumed by \citet{bienayme},
then $R_\mathrm{\nu}$ can be smaller than our estimate by up to a factor of two, falling well below 2 kpc.

The orientation of the velocity ellipsoid measured by the vertical gradient of $\overline{v_\mathrm{R} v_\mathrm{z}}$ has a minor impact on the radial scalelength. With $\eta\to 0$ (horizontal orientation), the scalelength $R_\mathrm{E}$ would increase by less than 20\%. The new St\"omberg relation (see Eq. \ref{new-stromberg}) shows that a re-determination of the velocity ellipsoid has, in general, a small effect on the determination of the LSR and the radial scalelengths.

Based on the large data sample of RAVE dwarfs, we have demonstrated that the Jeans equation is problematic for the determination of both the LSR (i.e. the tangential peculiar motion of the Sun $V_\mathrm{\sun}$) and the radial scalelengths of the stellar populations simultaneously. The extrapolation to the asymmetric drift value at $\sigma_\mathrm{R}=0$ depends sensitively on the sample selection and on additional assumptions. On the other hand, the Jeans equation provides a sensitive tool to test Milky Way models on their dynamical consistency. The LSR value cannot be adjusted independently because any variation has a large impact on the radial scalelengths of all stellar populations in dynamical equilibrium. Dynamically, the radial scalelength $R_\mathrm{E}$  of the radial energy density $\nu\sigma_\mathrm{R}^2$ is relevant, and its split into the scalelengths of the density and of the velocity dispersion needs further information, such as the vertical thickness in combination with the shape of the velocity dispersion ellipsoid.

\section*{Acknowledgements}

OG acknowledges funding by International Max Planck Research School for Astronomy \& Cosmic Physics at the
University of Heidelberg.

This work was supported by Sonderforschungsbereich SFB 881 `The Milky Way
System' (subproject A6) of the German Research Foundation (DFG).

Funding for RAVE has been provided by: the Australian Astronomical
Observatory; the Leibniz-Institut f\"{u}r Astrophysik Potsdam (AIP); the
Australian National University; the Australian Research Council; the
French National Research Agency; the German Research Foundation (SPP 1177 and SFB 881);
the European Research Council (ERC-StG 240271 Galactica); the Istituto Nazionale di
Astrofisica at Padova; The Johns Hopkins University; the National Science Foundation of
the USA (AST-0908326); the W. M. Keck foundation; the Macquarie University; the
Netherlands Research School for Astronomy; the Natural Sciences and Engineering Research
Council of Canada; the Slovenian Research Agency; the Swiss National Science Foundation;
the Science \& Technology Facilities Council of the UK; Opticon; Strasbourg Observatory;
and the Universities of Groningen, Heidelberg, and Sydney. The RAVE web site is at
http://www.rave-survey.org.

Funding for SDSS-I and SDSS-II has been provided by the Alfred P. Sloan Foundation, the Participating Institutions,
the National Science Foundation, the U.S. Department of Energy, the National Aeronautics and Space Administration,
the Japanese Monbukagakusho, the Max Planck Society, and the
Higher Education Funding Council for England. The SDSS Web Site is http://www.sdss.org/.

The authors are very grateful to Young-Sun Lee and Timothy C. Beers for providing their SEGUE data sample for our analysis and for fruitful discussions,
as well as to Hartmuth Jahrei{\ss} for providing us with results of his analysis of the local stellar samples.
We thank the anonymous referee for valuable comments and the language editor for improving the English language.



\end{document}